\documentclass{article}
\usepackage{spconf,amsmath,graphicx}
\usepackage{subfigure}
\usepackage[hidelinks]{hyperref}
\usepackage{multicol, multirow}
\usepackage{xcolor}
\usepackage[numbers, sort]{natbib}
\definecolor{Mulberry}{rgb}{0.77,0.29,0.55}
\definecolor{CadmiumOrange}{rgb}{0.93,0.53, 0.18}
\definecolor{ForestGreen}{rgb}{0.13, 0.55, 0.13}
\definecolor{WildStrawberry}{rgb}{0.5, 0.7, 0.2}

\title{DJCM: A Deep Joint Cascade Model for Singing Voice Separation and Vocal Pitch Estimation}
\name{Haojie Wei$^{1}$, Xueke Cao$^{2}$, Wenbo Xu$^{1}$, Tangpeng Dan$^1$, Yueguo Chen$^{1,*}$~\thanks{$^{*}$ Corresponding Author}}
\address{$^1$School of Information, Renmin University of China, Beijing, China\\
$^2$Institute of Information Science, Beijing Jiaotong University, Beijing, China}
\begin{document}
%
\maketitle
\begin{abstract}
Singing voice separation and vocal pitch estimation are pivotal tasks in music information retrieval. 
Existing methods for simultaneous extraction of clean vocals and vocal pitches can be classified into two categories: pipeline methods and naive joint learning methods.
However, the efficacy of these methods is limited by the following problems:
On the one hand, pipeline methods train models for each task independently, resulting a mismatch between the data distributions at the training and testing time.
On the other hand, naive joint learning methods simply add the losses of both tasks, possibly leading to a misalignment between the distinct objectives of each task.
To solve these problems, we propose a \textbf{D}eep \textbf{J}oint \textbf{C}ascade \textbf{M}odel (DJCM) for singing voice separation and vocal pitch estimation.
DJCM employs a novel joint cascade model structure to concurrently train both tasks. 
Moreover, task-specific weights are used to align different objectives of both tasks.
Experimental results show that DJCM achieves state-of-the-art performance on both tasks, with great improvements of 0.45 in terms of Signal-to-Distortion Ratio (SDR) for singing voice separation and 2.86\% in terms of Overall Accuracy (OA) for vocal pitch estimation. 
Furthermore, extensive ablation studies validate the effectiveness of each design of our proposed model.
The code of DJCM is available at \url{https://github.com/Dream-High/DJCM}.
\end{abstract}
\section{Introduction}\label{sec:intro}
Singing voice separation (SVS) and vocal pitch estimation (VPE) are two fundamental and highly related tasks in music information retrieval (MIR). 
The SVS task involves extracting clean vocals from mixture music, enabling downstream applications such as lyrics extraction~\cite{fan2016singing}, voice detection~\cite{jdc2019} and vocal pitch estimation~\cite{gao2021vocal}. 
Concurrently, the VPE task aims to identify vocal pitches from either clean vocals or mixture music.
The precise identification of vocal pitches holds paramount importance in various MIR applications, including content-based music recommendation, query/search by singing~\cite{ijcai2023p544} and score-informed singing voice separation~\cite{scoresvs2014}.

Most of existing methods have primarily focused on either SVS or VPE independently.
For example, methods like U-Net~\cite{svsunet2017}, Spleeter~\cite{spleeter2020} and ResUNetDecouple+~\cite{kong2021decoupling} predict vocals in the frequency domain. 
Conversely, Demucs~\cite{defossez2019music}, Wave-U-Net~\cite{wavunet0-2018} and its follow-ups~\cite{wavunet1-2018, wavunet-2020} employ time domain features for SVS. 
Other methods, such as KUIELAB-MDX-Net~\cite{kim2021kuielab}, Hybrid Demucs~\cite{defossez2021hybrid} and HT Demucs~\cite{rouard2022hybrid}, combine features from both frequency and time domains to enhance the performance of SVS. 
Meanwhile, VPE is addressed by heuristic-based methods (e.g., YIN~\cite{de2002yin}, SWIPE~\cite{swipe2008}, and pYIN~\cite{mauch2014pyin}) that utilize candidate-generating functions, 
and data-driven methods (e.g., CREPE~\cite{kim2018crepe}, DeepF0~\cite{deepf0-2021}, and HARMOF0~\cite{harmof0_2022}) employing supervised neural network training on clean vocals.
Although these methods have achieved good results on SVS or VPE, these methods focus solely on their respective tasks.

Based on the above methods for SVS and VPE, the most straightforward approach for simultaneous extraction of clean vocals and vocal pitches from mixture music is the pipeline method.
Pipeline methods (e.g., DNN+UPDUDP~\cite{dnn2016} and HR-ED~\cite{gao2021vocal}) usually train models for SVS and VPE separately on different music datasets.
These methods firstly extract clean vocals from mixture music by using trained SVS models, followed by extracting vocal pitches from predicted vocals by using trained VPE models.
However, the improvement of VPE is limited since there is a mismatch between the data distributions at the training and testing time.

Besides the aforementioned pipeline methods, a few joint learning methods (e.g., HS-W$_{p}$~\cite{nakano2019joint} and Joint-UNet~\cite{joint2019unet}) have been proposed for the simultaneous SVS and VPE tasks.
However, HS-W$_{p}$ overlooks diverse joint model structures, while Joint-UNet~\cite{joint2019unet} primarily investigates various joint model structures grounded in the UNet architecture~\cite{svsunet2017}.
Additionally, both methods simply add the losses of two tasks, potentially leading to the misalignment of task objectives.
Thus, the performance of these methods for SVS and VPE is limited although they can extract clean vocals and vocal pitches from the mixture music at the same time.

\begin{figure*}[htbp]
\centerline{\includegraphics[width=0.9\textwidth]{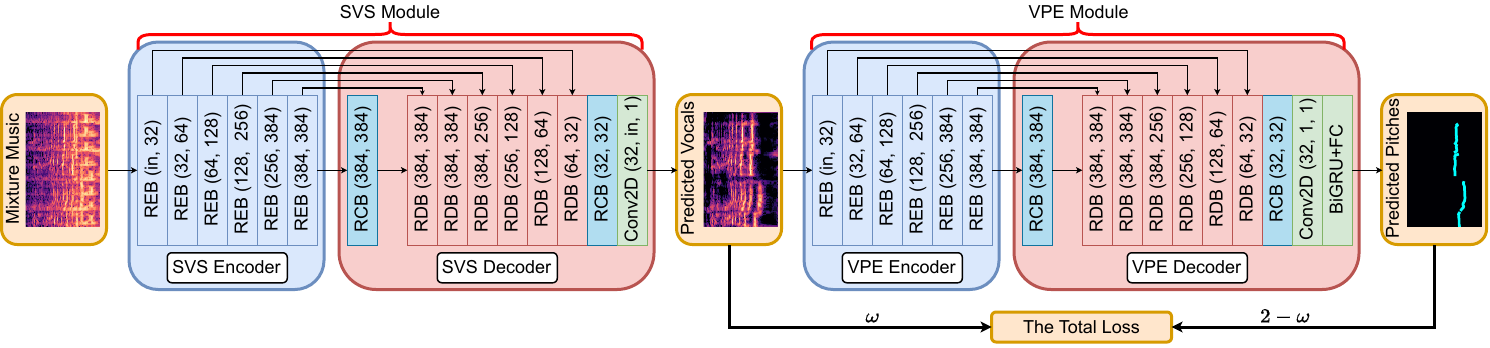}}
\caption{The overall structure of our proposed model DJCM.}
\label{fig:structure}
\end{figure*}

To address the above problems, we propose a \textbf{D}eep \textbf{J}oint \textbf{C}ascade \textbf{M}odel (DJCM) for SVS and VPE.
By comparing various model structures for joint learning, we design our deep joint cascade model for both tasks.
This model enables concurrent training and testing of SVS and VPE tasks.
Furthermore, we strategically assign task-specific weights to align different objectives of both tasks.
Besides, extensive experimental results demonstrate the effectiveness of our model.

\section{Method}\label{sec:method}
To solve the problems introduced in Section~\ref{sec:intro}, we propose a deep joint cascade model (DJCM) for singing voice separation and vocal pitch estimation.
The problem formulation, overall structure and the loss function are introduced in Section~\ref{sec:pb}, Section~\ref{sec:ma} and Section~\ref{sec:loss}, respectively.

\subsection{Problem Formulation}\label{sec:pb}
\textbf{Singing Voice Separation (SVS).} The SVS task involves extracting clear vocals from mixture music.
The mixture music can be represented as either the raw audio waveform $x$ or its corresponding spectrogram ${X}_{T\times F}$, where $T$ is the number of audio frames and $F$ is the number of frequency bins. 
The clean vocals is the output of SVS, which can be represented as either the raw audio waveform $v$ or its corresponding spectrogram ${V}_{T\times F}$.
Thus, the SVS task can be formulated as $\mathcal{F}_{svs}: x ({X}_{T\times F}) \to v$.

\noindent\textbf{Vocal Pitch Estimation (VPE).} The VPE task aims to extract vocal pitches from clean vocals or mixture music.
Following CREPE~\cite{kim2018crepe}, each vocal pitch is typically represented as a 360-dimensional one-hot vector $y$, and the output of VPE is a sequence of pitch vectors $Y_{T\times 360}$. 
Besides, calculating vocal pitches from $y$ is the same as which in CREPE~\cite{kim2018crepe}.
Thus, the VPE task from clean vocals can be formulated as $\mathcal{F}_{vpe}: v ({V}_{T\times F}) \to Y_{T\times 360}$.
And the VPE task from mixture music can be formulated as $\mathcal{F}_{vpe}: x ({X}_{T\times F}) \to Y_{T\times 360}$.

\subsection{Model Architecture}\label{sec:ma}
The overall deep joint cascade model (DJCM) structure consists of singing voice separation module (SVS Module) and vocal pitch estimation module (VPE Module), as shown in Fig.~\ref{fig:structure}. 
The SVS Module processes the mixture music spectrogram (${X}_{T\times F}$) and employs an encoder with 6 REBs, while the VPE Module processes predicted vocals spectrogram (${\hat{V}}_{T\times F}$) using a similar encoder.
The details of both modules are introduced as follows.

\begin{figure}[htbp]
    \centering
    \subfigure[]{
        \label{fig:2a}
        \centering
        \includegraphics[height=0.3\columnwidth]{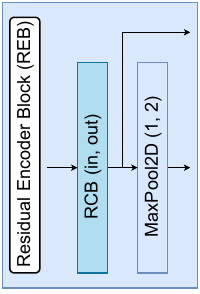}}
    \hspace{4mm}
    \subfigure[]{
        \label{fig:2b}
        \centering
        \includegraphics[height=0.3\columnwidth]{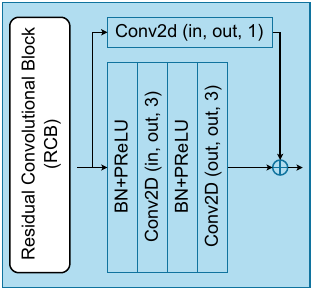}}
    \hspace{4mm}
    \subfigure[]{
        \label{fig:2c}
        \centering
        \includegraphics[height=0.3\columnwidth]{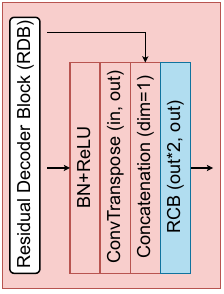}}
    \caption{(a) Residual Encoder Block (REB), (b) Residual Convolutional Block (RCB), (c) Residual Decoder Block (RDB)}
\end{figure}

The SVS Module takes the spectrogram of mixture music (${X}_{T\times F}$) as the input features. 
These features firstly pass through the SVS encoder, which consists of 6 residual encoder blocks (REBs). 
Each REB contains a residual convolutional block (RCB) following a max pooling layer with a kernel size 1 × 2 as shown in Fig.~\ref{fig:2a}.
The details of RCB are shown in Fig.~\ref{fig:2b}. 
There are 2 convolutional layers with $3\times3$ kernels. 
Before each convolutional layer, there is a BN layer and a PReLU function. 
Additionally, there is a shortcut connection between the input and the output of a RCB.

After the layers of SVS encoder, the SVS decoder extracts the spectrogram of clean vocals (${V}_{T\times F}$) from hidden features.
The bottom layer of SVS decoder is a RCB.
Then there are 6 residual decoder blocks (RDBs). 
Each RDB includes a transposed convolutional layer with a kernel size of $3\times3$ and stride of $1\times2$ to upsample feature maps, followed by a RCB, as shown in Fig.~\ref{fig:2c}.
The final components of SVS Decoder is a RCB and a convolutional layer with $1\times1$ kernel size.
At last, we use the iSTFT~\cite{torchlibrosa2019} to transform the spectrogram (${\hat{V}}_{T\times F}$) to the raw audio waveform ($\hat{v}$) of predicted vocals.

Then, the VPE Module takes the spectrogram of predicted vocals (${\hat{V}}_{T\times F}$) as the input features.
These features firstly pass through the VPE encoder, which has the same structure as the SVS encoder as shown in Fig.~\ref{fig:structure}.
Then the hidden features pass through the VPE decoder to get predicted vocal pitches ($\hat{y}$).
The VPE decoder has more layers of a bidirectional GRU and a fully connected sigmoid layer with 360 outputs compared with the SVS decoder.
At last, the final vocal pitches are calculated from predicted values $\hat{y}$ using the same way in CREPE~\cite{kim2018crepe}.

\subsection{The Loss of Joint Learning}\label{sec:loss}
When training our model, we employ the mean absolute error (MAE)~\cite{defossez2019music} loss for SVS and the weighted binary cross-entropy (BCE)~\cite{wei23b_interspeech} loss for VPE.

Then the loss function for SVS is defined as follows:
\begin{equation} 
\mathcal{L}_{svs}(v, \hat{v}) = \frac{1}{L}\sum_{i}^{L}|v_{i}-\hat{v_{i}}|
\end{equation}
where $L$ is the length of mixture music.
Additionally, the ground-truth for the pitches of each frame is represented by a 360-dimensional one-hot vector donated as $y$. 
Thus, an imbalance arises between the positive and negative categories when employing the BCE loss.
Based on this observation, we use the weighted binary cross entropy (BCE).
Then the loss function for VPE is defined as follows:
\begin{equation} 
\mathcal{L}_{vpe}(y, \hat{y}) = -\sum_{i}^{360} (\alpha y_{i} \log \hat{y_{i}}+(1-y_{i}) \log(1- \hat{y_{i}}))
\end{equation}
where we set the weight parameter $\alpha$ as 10 in this paper. 

The naive joint learning method involves the straightforward summation of the losses from SVS and VPE. 
However, this approach may lead to misalignment between the distinct objectives of these tasks.
Hence, we use a hyper-parameter $\omega$ to dynamically adjust the weights assigned to different tasks, thereby aligning different objectives of both tasks.
For consistency with the naive joint learning method, we align the tasks using a weight sum of 2.
Thus, the total loss for our joint learning method is defined as:
\begin{equation} 
\mathcal{L}_{total} = \omega\times\mathcal{L}_{svs} + (2-\omega)\times\mathcal{L}_{vpe}
\end{equation}
Notably, when $\omega$ is set to 1, this loss is the same as which in the naive joint learning method.

\section{Experimental Setup}
\subsection{Datasets} \label{sec:dataset}
To compare with previous methods fairly, we adopt the public dataset, MIR-1K~\cite{mir1k2010}.
It is a dataset designed for singing voice separation.
This dataset contains 1000 song clips with the music accompaniment and the singing voice recorded as left and right channels, respectively. 
Besides, the MIR-1K dataset has the labels of vocal pitches in semitone.
To the best of our known, MIR-1K is the only dataset available for evaluating both the SVS and VPE tasks.

\subsection{Evaluation Metrics}
The following evaluation metrics are used to evaluate the performance of singing voice separation and vocal pitch estimation, as described in previous studies~\cite{kong2021decoupling, mir1k2010, kim2018crepe, wei23b_interspeech}.
These metrics are computed using the mir\_eval~\cite{raffel2014mir_eval} library. 
The details are illustrated as follows:

\textbf{Signal-to-Distortion Ratio (SDR)}~\cite{kong2021decoupling} is used to measure the quality of the predicted vocals with respect to clean vocals. 
It is defined as $SDR(v, \hat{v}) = 10\times\log_{10}{\frac{||v||^2}{||\hat{v}-v||^2}}$, where $\hat{v}$ is the predicted vocals and $v$ is the clean vocals. 
A higher SDR indicates better separation results, and vice versa. Ideally, a perfect separation will lead to infinite SDR.
\textbf{Global Normalized Signal-to-Distortion Ratio (GNSDR)}~\cite{mir1k2010} is calculated as $GNSDR = \frac{\sum_{i=1}^{i=N}l_iNSDR(v, \hat{v}, x)}{\sum_{i=1}^{i=N}l_i}$,
where $i$ is the index of a song, $N$ is the total number of songs, $l_i$ is the length of the $ith$ song and $NSDR(v, \hat{v}, x)$ is the normalized SDR. 
The NSDR~\cite{mir1k2010} is defined as $NSDR(v, \hat{v}, x) = SDR(v, \hat{v}) - SDR(v, x)$, where $x$ is the mixture music. 
The NSDR is the improvement of SDR between the mixture music and predicted vocals.
\textbf{Raw Pitch Accuracy (RPA)}~\cite{kim2018crepe} computes the proportion of melody frames in the reference for which the predicted pitch is within $\pm50$ cents of the ground truth pitch.
\textbf{Raw Chroma Accuracy (RCA)}~\cite{kim2018crepe} 
measures raw pitch accuracy ignoring octave errors.
\textbf{Overall Accuracy (OA)}~\cite{raffel2014mir_eval} computes the proportion of all frames correctly estimated by the model, including whether non-melody frames where labeled by the model as non-melody.

\subsection{Implementation Details}\label{sec:im_details}
The raw audio is sampled at 16kHz and then transformed into a spectrogram by the Short-Time Fourier Transform (STFT). The hop length of the spectrogram is 320 (20ms) and the Hann window size is 2048. The torchlibrosa~\cite{torchlibrosa2019} is adopted to finish the above audio processing.
We use the Adam optimizer~\cite{kingma2014adam} to train DJCM and set the batch size as 16. The learning rate is initialized as 5e-4 and reduced by 0.95 of the previous learning rate every 5 epochs. Each training audio is divided into 2.56-second segments. 
We randomly spilt the MIR-1K dataset into training (80\%) set and testing (20\%) set.
Besides, all experiments are conducted on MIR-1K dataset.

\section{Experimental Results}
In this section, we present the experimental results of DJCM in the tasks of singing voice separation (SVS) and vocal pitch estimation (VPE).
We firstly compare our model with previous methods on MIR-1K dataset. 
Then we explore different model structures for joint learning of both tasks to show our model structure is effective. 
Lastly, we conduct ablation studies to analyze the effectiveness of our joint learning method.

\subsection{Comparison With Previous Methods}
To evaluate the superiority of our framework, we compare our model with previous methods on the MIR-1K dataset.
The baselines divided into joint learning methods and pipeline methods, which are trained and tested on the MIR-1K dataset.

\begin{table}[htbp]
\centering
\caption{Performance comparison for singing voice separation and vocal pitch estimation on MIR-1K.}
\label{tab:main results}
\resizebox{\columnwidth}{!}{
\begin{tabular}{l|cc|ccc}
\hline
\multirow{2}{*}{Methods} & \multicolumn{2}{c|}{SVS} &\multicolumn{3}{c}{VPE (\%)}         \\ \cline{2-6}
                   & SDR   & GNSDR & RPA & RCA &OA \\ \hline
\textbf{\textit{Previous Joint Learning Methods}}    &&&& \\                   
HS-W$_{p}$~\cite{nakano2019joint}           &9.80 &6.87 &85.04 &85.32&83.65  \\ 
S$\rightarrow$P$\rightarrow$S$\rightarrow$P~\cite{joint2019unet}              &11.70&8.72&86.62&86.94 & 85.41\\ \hline
\textbf{\textit{Pipeline Methods}} w/i CREPE~\cite{kim2018crepe}    &&&& \\ 
U-Net~\cite{svsunet2017}        &11.43&8.48&89.28&90.41      &87.56      \\ 
ResUNetDecouple+~\cite{kong2021decoupling}     &12.06&9.13&91.40&92.07   &89.97           \\ \hline
\textbf{\textit{Pipeline Methods}} w/i HARMOF0~\cite{harmof0_2022}    &&&& \\     
U-Net~\cite{svsunet2017}       &11.43&8.48&87.95&88.57&86.35          \\ 
ResUNetDecouple+~\cite{kong2021decoupling}   &12.06&9.13&90.21&90.61 &88.24  \\ \hline
DJCM    & \textbf{12.51} & \textbf{9.50}  & \textbf{93.72} & \textbf{93.96}   & \textbf{92.83}  \\ \hline
\end{tabular}
}
\end{table}
Our model demonstrates great effectiveness in both SVS and VPE tasks, achieving state-of-the-art performance, as summarized in Table~\ref{tab:main results}.
Compared to previous joint learning and pipeline methods, our model achieves great improvement on both tasks.
Specifically, DJCM outperforms previous best joint learning method (S$\rightarrow$P$\rightarrow$S$\rightarrow$P) by 0.81 in SDR, 7.10\% in RPA and 7.42\% in OA.
These results show the effectiveness of our model structure and joint learning method.
Furthermore, compared to best pipeline method (ResUNetDecouple+ with CREPE), DJCM improves 0.45 in SDR, 2.32\% in RPA and 2.86\% in OA.
This is because singing voice separation and vocal pitch estimation models trained separately in pipeline methods, and vocal pitch estimation models in pipeline methods are greatly affected by unseparated accompaniment.
While our model trains both tasks simultaneously, making VPE Module of DJCM is more robust with unseparated accompaniment in predicted vocals.

\subsection{Ablation Study for Model Structure}
\begin{figure}[htbp]
    \centering
    \subfigure[]{
        \label{fig:3a}
        \centering
        \includegraphics[height=0.3\columnwidth]{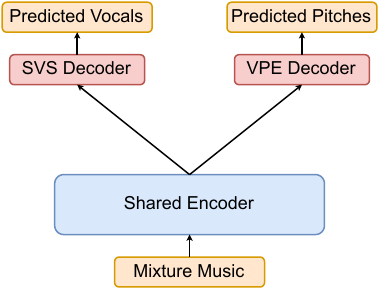}}
    \hspace{4mm}
    \subfigure[]{
        \label{fig:3b}
        \centering
        \includegraphics[height=0.3\columnwidth]{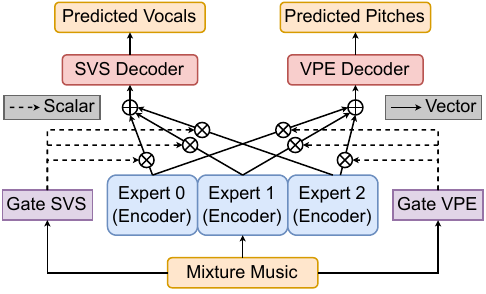}}
    \caption{(a) The Shared Bottom structure. (b) The MMOE structure, the number of experts can be adjusted.}
    \label{fig:3}
\end{figure}

To investigate the effectiveness of our joint cascade model structure for joint learning of SVS and VPE, we compare DJCM with two alternative structures: the Shared Bottom structure and the MMOE structure as shown in Fig.~\ref{fig:3}. 
The shared bottom structure (c.f. Fig.~\ref{fig:3a}) makes the SVS and VPE tasks sharing bottom layers as the shared encoder.
While the MMOE structure employs multiple ($n$) experts with gating mechanisms as the ensemble shared encoder to extract hidden features.
With $n$ experts, the MMOE structure can be represented as MMOE-$n$.
Besides, the single encoder and decoder structures in Fig.~\ref{fig:3} are the same as which in Fig.~\ref{fig:structure}.

As shown in Table~\ref{tab:structure results}, DJCM achieves the best performance among all model structures.
\begin{table}[htbp]
\centering
\caption{Performance with different model structures on MIR-1K.}
\label{tab:structure results}
\resizebox{0.9\columnwidth}{!}{
\begin{tabular}{l|cc|ccc}
\hline
\multirow{2}{*}{Model Structures} & \multicolumn{2}{c|}{SVS} &\multicolumn{3}{c}{VPE (\%)}         \\ \cline{2-6}
                   & SDR   & GNSDR & RPA & RCA &OA \\ \hline    
\textbf{\textit{Single Model}}    &&&& \\   
SVS            &12.16    & 9.15   & ——    & —— & ——   \\ 
VPE                & ——    & ——    &92.64   & 92.92 &91.10  \\ \hline
\textbf{\textit{Joint Model}}    &&&& \\   
Shared Bottom     & 12.42  & 9.41 & 92.89   & 93.19   & 91.66          \\ 
MMOE-2   & 11.87 & 8.86  & 92.54   & 92.86   & 92.02          \\ 
MMOE-3         & 12.43 & 9.42  & 93.58   & 93.96   & 90.16            \\ \hline
DJCM    & \textbf{12.51} & \textbf{9.50}  & \textbf{93.72} & \textbf{93.96}   & \textbf{92.83}  \\ \hline
\end{tabular}
}
\end{table}
Compared to the single SVS model, DJCM improves 0.35 on SDR. 
And compared to the single VPE model, DJCM improves 1.08\% on RPA and 1.73\% on OA.
These results show joint learning of both tasks can make both tasks benefit for each other.
Moreover, DJCM gets the better performance than any other joint model structures. 
Specifically, DJCM outperforms the Shared Bottom structure by 0.83\% in RPA and 1.17\% in OA. 
And compared to MMOE-2, our model improves 0.64 in SDR, 1.18\% in RPA and 0.81\% in OA.
These results show that our joint cascade model structure is best among these joint model structures and joint learning of both tasks can make the performance of both tasks achieving a great improvement.

\subsection{Ablation Study for Joint Learning}
To explore the effectiveness of the joint learning method, we set different values of $\omega$ in the total loss of joint learning.
The value of $\omega$ ranges from 0.2 to 1.8 in steps of 0.2.
The results for different $\omega$ values are summarized in Table~\ref{tab:omega results}.
\begin{table}[htbp]
\centering
\caption{Performance with different values of $\omega$ on MIR-1K.}
\label{tab:omega results}
\resizebox{0.75\columnwidth}{!}{
\begin{tabular}{l|cc|ccc}
\hline
\multirow{2}{*}{$\omega$} & \multicolumn{2}{c|}{SVS} &\multicolumn{3}{c}{VPE (\%)}         \\ \cline{2-6}
                   & SDR   & GNSDR & RPA & RCA &OA \\ \hline    
0.2         & 10.27 & 7.26  & 93.67   & 93.95   & 92.56            \\ \hline
0.4         & 11.26 & 8.25  & 93.14   & 93.35   & 92.77            \\ \hline
0.6         & 11.05 & 8.04  & \textbf{93.96}   & \textbf{94.22}   & 91.59            \\ \hline
0.8         & 11.33 & 8.32  & 93.71   & 93.92   & 92.18            \\ \hline
1.0         & 11.98 & 8.97  & 92.51   & 92.74   & 91.56            \\ \hline
1.2         & 11.58 & 8.57  & 93.84   & 94.04   & 92.10            \\ \hline
1.4         & 12.38 & 9.36  & 93.45   & 93.70   & 92.94            \\ \hline
1.6         & 12.01 & 9.00  & 93.61   & 93.79   & \textbf{92.94}            \\ \hline
1.8    & \textbf{12.51} & \textbf{9.50}  & 93.72 &93.96   & 92.83  \\ \hline
\end{tabular}
}
\end{table}

In this experiment, reducing the importance of the separation objective implies increasing the importance of VPE, making the whole model closer to the end-to-end methods.
When $\omega$ changes from 0.2 to 1.0, the RPA decreases by 1.16\%, and when $\omega$ changes from 1.0 to 1.8, the RPA increases by 1.21\%. 
These results indicate that the VPE performance is significantly affected by the value of weights.

Moreover, the best performance on pitch transcription is achieved with a sub-optimal vocal separation module. 
This is because, when $\omega$ is set to 0.6, the joint learning method emphasizes VPE, leading to the best VPE performance. 
When $\omega$ is larger than 1.0, SVS becomes more important, improving SVS performance and consequently enhancing VPE performance due to the cascade model structure.

Furthermore, when $\omega$ is set to 1, DJCM simply adds the losses of both tasks, similar to previous naive joint learning methods. 
Compared to the single model of singing voice separation (c.f. SVS in Table~\ref{tab:structure results}), DJCM decreases by 0.18 in SDR. 
This result shows that the different objectives of both tasks in naive joint learning are misaligned.
When $\omega$ is set to 0.6, DJCM achieves the best performance in vocal pitch estimation, improving RPA by 1.22\% compared to naive joint learning.
When $\omega$ is set to 1.8, our model achieves the best performance in singing voice separation, improving SDR by 0.53 compared to naive joint learning.
These results show that our joint learning method can align the different objectives in joint learning, making our model effective in joint learning of both tasks.  

\section{Conclusion}
In this paper, we propose a deep joint cascade model (DJCM) to extract clean vocals and vocal pitches at the same time. 
DJCM can extract singing voice and vocal pitches at the same time.
The experimental results show that DJCM is effective for joint learning of singing voice separation and vocal pitch estimation, since it achieves state-of-the-art performance on the MIR-1K dataset. 
In the future, there are two directions can be further explored. 
They are more suitable joint learning model structures and more effective multi-task optimization methods for joint cascade model.

\section{Acknowledgements}
This work is supported by the National Science Foundation of China under the grant 62272466, and Public Computing Cloud, Renmin University of China.
\bibliographystyle{IEEEbib}
\bibliography{strings,refs}

\end{document}